\documentclass{article}

\usepackage{arxiv}

\usepackage[utf8]{inputenc} % allow utf-8 input
\usepackage[T1]{fontenc}    % use 8-bit T1 fonts
\usepackage{hyperref}       % hyperlinks
\usepackage{url}            % simple URL typesetting
\usepackage{booktabs}       % professional-quality tables
\usepackage{amsfonts}       % blackboard math symbols
\usepackage{nicefrac}       % compact symbols for 1/2, etc.
\usepackage{microtype}      % microtypography
\usepackage{lipsum}		% Can be removed after putting your text content
\usepackage{graphicx}
\usepackage{natbib}
\usepackage{multirow}
\usepackage{doi}
\usepackage{amsmath,amssymb,amsfonts}
\usepackage{xcolor}

\usepackage{hyperref}

\definecolor{newcolor}{rgb}{.8,.349,.1}
\usepackage{subcaption} % Load the subcaption package

\title{DSAM: A Deep Learning Framework for Analyzing Temporal and Spatial Dynamics in
Brain Networks}

\author{\thanks{Corresponding Author: Bishal Thapaliya, bthapaliya16@gmail.com.}\, Bishal Thapaliya \textsuperscript{1, 2} \quad Robyn Miller \textsuperscript{5}  \quad Jiayu Chen \textsuperscript{1, 2} \quad Yu-Ping Wang \textsuperscript{4} \quad Esra Akbas \textsuperscript{1} \quad Ram Sapkota \textsuperscript{1, 2} \\\quad \textbf{Bhaskar Ray} \textsuperscript{1, 2} \quad \textbf{Pranav Suresh \textsuperscript{1, 2}} \quad \textbf{Santosh Ghimire \textsuperscript{6}} \quad \textbf{ Vince Calhoun \textsuperscript{1, 2, 3}} \quad \textbf{Jingyu Liu \textsuperscript{1,2}}\\
\textsuperscript{1}Georgia State University  \quad  \textsuperscript{2}TReNDs Center \quad \textsuperscript{3}Georgia Institute of Technology \\\quad \textsuperscript{4}Tulane University \quad \textsuperscript{5}Los Alamos National Laboratory \quad \textsuperscript{6}Tribhuvan University
\\}

\renewcommand{\shorttitle}{\textit{Dynamic Spatio-Temporal Attention Model (DSAM )} }

\begin{document}
\maketitle

\newcommand{\technique}{\textit{DSAM }}
\maketitle

\begin{abstract}
Resting-state functional magnetic resonance imaging (rs-fMRI) is a noninvasive technique pivotal for understanding human neural mechanisms of intricate cognitive processes. Most rs-fMRI studies compute a single static functional connectivity matrix across brain regions of interest, or dynamic functional connectivity matrices with a sliding window approach. These approaches are at risk of oversimplifying brain dynamics and lack proper consideration of the goal at hand. While deep learning has gained substantial popularity for modeling complex relational data, its application to uncovering the spatiotemporal dynamics of the brain is still limited. In this study we propose a novel interpretable deep learning framework that learns goal-specific functional connectivity matrix directly from time series and employs a specialized graph neural network for the final classification. Our model, \technique, leverages temporal causal convolutional networks to capture the temporal dynamics in both low- and high-level feature representations, a temporal attention unit to identify important time points, a self-attention unit to construct the goal-specific connectivity matrix, and a novel variant of graph neural network to capture the spatial dynamics for downstream classification. To validate our approach, we conducted experiments on the Human Connectome Project dataset with 1075 samples to build and interpret the model for the classification of sex group, and the Adolescent Brain Cognitive Development Dataset with 8520 samples for independent testing. Compared our proposed framework with other state-of-art models, results suggested this novel approach goes beyond the assumption of a fixed connectivity matrix, and provides evidence of goal-specific brain connectivity patterns, which opens up potential to gain deeper insights into how the human brain adapts its functional connectivity specific to the task at hand. Our implementation can be found on \url{https://github.com/bishalth01/DSAM}.
\end{abstract}

% keywords can be removed

\section{Introduction}

Resting-state functional magnetic resonance imaging (rs-fMRI) has emerged as a noninvasive technique pivotal for unraveling the intricate function of the human brain. This imaging modality has not only revolutionized our understanding of brain function but has also showcased its potential as a diagnostic tool for assessing brain disorders \cite{Fornito2015}.  By capturing intrinsic neural fluctuations during periods of rest, rs-fMRI enables researchers to delve into the dynamic interplay between brain regions, offering a gateway to comprehend functional connectivity patterns. The importance of understanding brain dynamics and related disorders has prompted significant efforts from fields such as traditional machine learning \cite{Britney2021, Ray2023, 2021EnvGenetics, Thapaliya2023} and especially deep learning \cite{Heinsfeld2018, RamMultiModal, Pranav}.

The analysis of rs-fMRI data relies on graph-theoretical metrics to abstract complex brain networks \cite{Azevedo2022}. Dimensionality reduction is a crucial step due to the high-dimensional nature of the data. This can be achieved through (1) temporal dimension reduction by constructing static connectivity matrices; (2) spatial dimension reduction by aggregating voxel-level signals into a predefined region of interests (ROIs) \cite{Wang2019}, and (3) combined temporal and spatial reduction methods like group independent component analysis \cite{Beckmann2005}. These steps help handle the large data volume and address the signal-to-noise ratio challenges \cite{Smith2018}. However, such reduction methods risk losing valuable information. For example, collapsing the temporal dimension oversimplifies brain dynamics, treating it as static, contrary to emerging evidence of continuous connectivity evolution over time \cite{AvenaKoenigsberger2017, Liao2017}. Additionally, common association measures often rely on linear models, despite the known nonlinear characteristics of brain signals and interactions \cite{Duggento2018, Goelman2018}.

A better understanding of brain connectivity variations necessitates the mathematical models capable of identifying brain disorder-specific irregularities \cite{Arslan, Ma}. Existing methods often compute functional connectivity (FC) matrices independently of the prediction task, limiting their adaptability. A notable attempt by \cite{Kim2021} computes FC based on learned data representations, yet it lacks adaptability in connectivity estimation. We propose using deep learning (DL) models with learnable weights to gauge goal-specific connectivity matrices.  This approach bridges connectivity analysis and predictive modeling, enabling us to uncover nuanced FC patterns in association with brain function and disorders \cite{Mahmood2022}.

Furthermore, FC estimates often follow static or dynamic computation approaches. Static connectivity assumes stationary brain activities,  while dynamic FC, on the other hand, reveals recurring patterns  that evade static approaches \cite{Allen2012}. Using a static graph-based method to understand dynamic systems can lead to reduced classification performance, as demonstrated by \cite{Xu2020}. Notably, \cite{Kipf} demonstrates improved outcomes by dynamically re-evaluating the learned static graph during testing, emphasizing the importance of leveraging dynamic nature of brain function. Modeling dynamic connectivity is therefore crucial \cite{Yaesoubi2018}. Yet, current methods frequently employs predefined sliding window techniques with window sizes and strides \cite{Armstrong2016,Damaraju2014}. In this study, we propose to use deep neural networks to present spatial and temporal dynamics and to build a goal-specific FC matrix.  This approach addresses the limitations of static methods and predefined sliding window techniques, offering insights into the dynamic nature of brain function.

Graph neural networks (GNN) are the most frequently used DL methods to study FC matrix. Most GNNs learn embeddings of nodes by just considering their neighborhood throughout the whole graph. However, this could be problematic with brain connectome due to the sub-network nature of the brain. Recently, BrainGNN \cite{Li2021} proposed a new GNN architecture that tackled this limitation by proposing a clustering-based embedding method in the graph convolutional layer, which allows nodes to consider their cluster assignment (representing different regions of the brain) while learning embeddings. Similarly, BrainRGIN \cite{BrainRGIN} proposed an isomorphism-based clustered convolution with attention-based readouts to improve the prediction power of GNNs. However, the major limitation of BrainGNN and BrainRGIN is that it did not capture the temporal dynamics of the brain.  
Beyond static FC GNN models,  FBNetGen (\cite{FBNetGen}) delves into the trainable generation of brain networks while investigating their interpretability for downstream tasks, utilizing a combination of FC features and learned feature representations at the final linear layer for the prediction. STAGIN (\cite{STAGIN}) employs GNNs with spatio-temporal attention to model dynamic brain networks extracted from fMRI data using dynamic FCs as input to the model. Graph Transformer \cite{GraphTransformer}  is a recent transformer-based method that uses static FC as features and applies multi-head attention to perform downstream tasks. (\cite{Azevedo2022}) introduces a slightly different idea of using raw rsfMRI timepoints as features instead of FC, and uses temporal blocks including (temporal) convolutional networks (TCN) or Long Short Term Memory (LSTM) to capture temporal information, and graph block to capture spatial information. A much more recent model, Brain Network Transformer (BNT) (\cite{BrainNetworkTransformer}) effectively used transformer-scaled dot product attention and introduced an orthonormal clustering readout function that leads to cluster-aware node embeddings. On the other hand, IBGNN (\cite{IBGNN}) initializes a mask based on edges and refines it with sparse control during training to identify top connections and regions for a specific task, aiming to create a task-specific sub-network for interpretation. Similarly, DICE \cite{Mahmood2022} proposed a novel idea of directed task-based connectivity matrix generation using temporal attention and self-attention. It is noteworthy that only a handful of papers like \cite{Mahmood2022, Azevedo2022} make use of only rsfMRI timepoints without making use of correlation based FC matrix.

In light of these challenges and opportunities, our paper proposes a new built-in interpretable model, Dynamic Spatio-Temporal Attention Model (\technique), that leverages the power of TCNs, self-attention, and GNNs to unfold the intricate spatiotemporal dynamics of brain directly from rsfMRI timepoints.  Through this approach, we strive to overcome the limitations of conventional methodologies and pave the way for a deeper comprehension of how the human brain learns and adapts its functional connectivity for various complex conditions. We want to point out that although we validate the model via sex classification, it can be used to predict/classify various phenotypes ranging from age, intelligence, and cognitive processes.

Overall, the main contribution of the paper lies in the built-in interpretability and end-to-end training, including (a) \textbf{temporal feature extraction directly from raw time series of the brain activity using a novel multilevel adaptation of TCNs,} (b) \textbf{shared temporal attention mechanisms to select informative time points,} (c) \textbf{node-node self-attention to build goal-specific brain connectivity matrix,} (d) \textbf{ROI-Aware GNNs for the spatial dynamics of the brain}.

The remainder of this paper is organized as follows: Section \ref{MethodsSection} describes our data and preprocessing step, Our \technique model is also
presented in Section \ref{MethodsSection}. Experiments and results are discussed
in Sections \ref{sec:replication_section} and \ref{sec:result}. Finally, Section \ref{sec:dic_con} concludes the paper with a discussion.

\section{Materials and Methods}

\label{MethodsSection}
\subsection{Data and Preprocessing }

\subsubsection{Human Connectome Project (HCP) young adult data}

 HCP rsfMRI data \cite{VanEssen2013} is first minimally pre-processed following the pipeline described in \cite{Glasser2013}. The preprocessing includes gradient distortion correction, motion correction, and field map preprocessing, followed by registration to T1 weighted image. The registered EPI image is then normalized to the standard MNI152 space. To reduce noise from the data, FIX-ICA based denoising is applied \cite{Griffanti2014, SalimiKhorshidi2014}. To minimize the effects of head motion, subject scans with framewise displacement (FD) over 0.3mm at any time of the scan were discarded. The FD is computed with fsl motion outliers function of the FSL \cite{Jenkinson2012}. There are 152 discarded scans after filtering out with the FD, and 1075 scans are left. For all experiments, the scans from the first run of HCP subjects released under S1200 are used. We use \cite{Schaefer2017} atlas for brain parcellation, with 100 regions. For each region, the average value is computed for all the voxels falling inside a region, thus resulting in a single time series for each region. After dividing data into regions, each time series is standardized by their score having zero mean and unit variance.

\subsubsection{Adolescent Brain Cognitive
Development (ABCD) Data}

ABCD is a large ongoing study following youths from age 9-10 into late adolescence to understand factors that increase
the risk of physical and mental health problems. Participants were recruited from 21 sites across the US to represent various demographic variables. Data used in this study are from 8520 children aged 9–10 at baseline, including resting state fMRI image and fluid intelligence, crystallized intelligence, and total composite scores. Data are splited into training (n = 5964), validation (n = 1278), and test (n = 1278) subsets. There are 4430 male and 4089 female subjects in this study.

We conduct preprocessing on the raw resting-state fMRI data utilizing a combination of the FMRIB Software Library (FSL) v6.0 toolbox and Statistical Parametric Mapping (SPM) 12 toolbox within the MATLAB 2019b environment. The preprocessing encompasses several key steps, namely: 1) correction for rigid body motion; 2) distortion correction; 3) removal of dummy scans; 4) standardization to the Montreal Neurological Institute (MNI) space; and 5) application of a 6 mm Gaussian kernel for smoothing. Similar to HCP dataset, we use \cite{Schaefer2017} atlas for brain parcellation, with 100 regions.

\subsection{The overview framework for DSAM }

\begin{figure*}
\includegraphics[width=\linewidth]{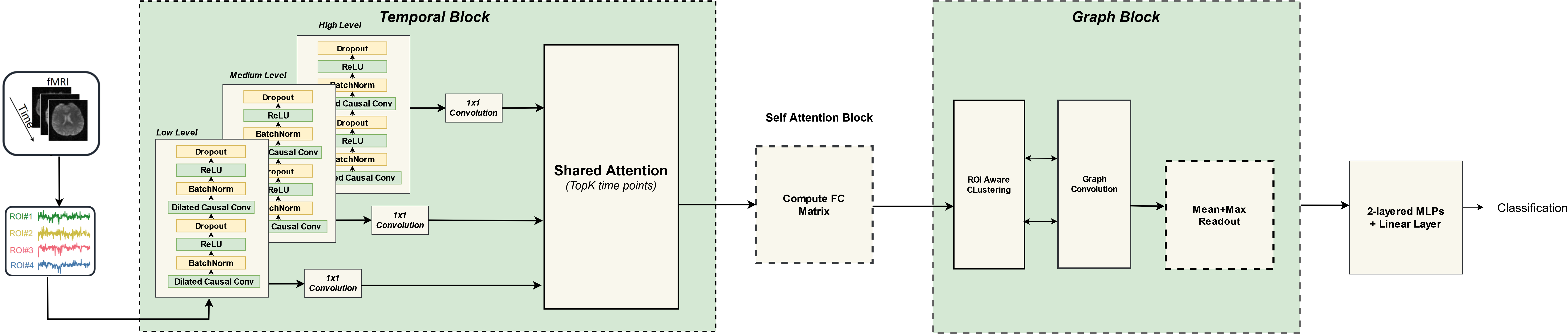}
\caption{  Overall Architecture of \technique.  3 TCN blocks are used to extract the temporal features with different levels of abstraction (low level, medium level, and high level). Temporal attention uses a shared multi head attention module to filter the important time points. Self Attention Block is used to learn the directed FNC matrix, which is used as an input to the Graph Block, followed by Graph Readout and a fully connected layer for classification. }

\label{DSAM_Architecture}
\end{figure*}

The Dynamic Spatio-Temporal Attention Model (DSAM) as presented in Figure \ref{DSAM_Architecture} is designed to effectively capture the intricate spatio-temporal dynamics inherent in complex systems such as brain networks. The DSAM architecture is comprised of several key components that work cohesively to extract informative features from spatio-temporal data and perform classification tasks. It begins with a Temporal Convolutional Network (TCN), capturing temporal dependencies in time series data. The TCN architecture consists of multiple blocks, each aimed at extracting different levels of feature from the input rsfMRI data. Following the TCN, the model incorporates a Temporal Attention Block, utilizing a Transformer-based multi-head attention mechanism, selectively considering informative temporal features in critical time points. Subsequently, the model employs a Self-Attention Block to calculate a learned connectivity matrix across nodes, representing the spatial network as input to the graph neural network model.  The graph structure is constructed either dynamically or statically by  Pearson correlation of the output from the Temporal Attention Block or original time series. The learned connectivity matrix, is passed as an input to the Relational Graph Isomorphism Network (RGIN) model to capture the spatial dynamics of the data. Finally, the output from RGIN is fed into a linear model for the classification task, leveraging the learned features and interactions within the graph structure. The architecture is further clearly elucidated in the subsections below.

\subsection{Temporal Convolutional Networks}
Temporal Convolutional Networks form the first component of our spatio-temporal model and are responsible for extracting temporal features from the 1D time-series data of each brain node. It is well documented that TCNs (\cite{Bai2018AnEE}) can perform equally well or sometimes even better than Recurrent Neural Networks (RNNs) and LSTMs for many kinds of sequential data.  Some advantages of the convolutional operator are, for instance, (1) low memory requirement for long input sequences, especially compared to LSTMs and Gated Recurrent Units (GRUs), (2) better parallelization because a TCN layer is processed as a whole instead of sequentially as in RNNs, and (3) easier to train (e.g., it is known that LSTM training can commonly encounter issues with vanishing gradients). TCNs are built on dilated causal convolutions, which are unique 1D filters whose receptive field size quickly grows across the temporal dimension of the input as the network's depth increases. To maintain temporal order, the padding of the convolution is "causal" in the sense that output at a particular time step is convolved exclusively with elements from earlier time steps from the previous layers.

Formally, given a single ROI time series \begin{math} t_i \in R^T \end{math} , i is the index of  node, T is the number of time  points, and a filter \begin{math} f \in R^K \end{math}, K is the kernel size, the dilated casual convolution operation with \begin{math} f \end{math} at time \begin{math} t \end{math} is represented as:

\begin{equation}
    (t_i * f)(t) = \sum_{s=0}^{K-1} f(s) \times t_i(t- d * s),
\end{equation} 
where \begin{math} d \end{math} is the dilation factor. When d = 1, a dilated convolution reduces to a regular convolution. Using larger dilation enables an output at the top level to effectively represent a wider receptive field of the input.\\

We propose a new multi-level adaptation of TCNs that includes three TCN blocks, and each block consists of two temporary convolution layers with multiple channels/kernels, batch normalization, and a dropout.  Similar to \cite{Bai2018AnEE}, we use batch normalization instead of weight normalization because of its stable performance during training.  Despite employing the same kernel size across all blocks, the dilation levels progressively increase along with the TCN blocks, as in Equation 2. 
 
\begin{equation}
    \tilde{T}_{nb} = TCN_{3}(TCN_{2}(TCN_1(t_{i}, d=1), d=2), d=4)
\end{equation}

where  $ \tilde{T}_{nb} \in \mathbb{R}^{N \times T}$ and N is the number of nodes. The three TCN blocks present three levels of features (high-level features, medium-level features, and low-level features) from the time series. At each of the three blocks, we employ distinct 1D convolutions across channels to obtain a single channel representation for that block.

\subsubsection{Shared temporal attention block}

To identify crucial time points, we introduce a \textit{shared attention } mechanism applied across the features extracted from different levels of TCN. This single attention mechanism filters important time points independently across different blocks while sharing weights. This design choice enforces the attention network to select a similar set of time points across all blocks, enabling a coherent selection of temporal patterns. Subsequently, we apply a top-k approach to filter out unimportant time points.

Initially, the outputs from three TCN blocks, $\tilde{T}_{1}, \tilde{T}_{2}$ and $\tilde{T}_{3}$ represented as  $\tilde{T}_{nb} \in \mathbb{R}^{(N \times T) } $across all nodes are individually fed to a \textit{shared attention} block to compute the individual attention scores as demonstrated below:

\begin{gather}
K_T = \tilde{T}_{nb} W_K, \quad Q_T = \tilde{T}_{nb} W_Q, \quad V_T = \tilde{T}_{nb} W_V \\
\text{AttentionMatrix}(Q_T, K_T) = \text{softmax}\left(\frac{Q_TK_T^T}{\sqrt{d_{k_T}}}\right) \\
\text{Attention}(Q_T, K_T, V_T) = \left(\text{AttentionMatrix}\right) V_T 
\end{gather}
where, $\textit{AttentionMatrix} \in \mathbb{R}^{T \times T}$, $Attention \in \mathbb{R}^{N \times T}$, %$m$ is the number of attention heads, % 
and $W_Q , W_K$, $W_V$ are the parameters of the attention block. We implemented multi-head attention(MHA) for this module, followed by  \textit{topK} approach to extract the important time points based on \textit{topK} threshold $th$ within each block. 

\begin{gather}
\hat{T}_{nb} = \textit{topK}\left(\text{MHA}(\tilde{T}_{nb}), th\right) \quad \text{for } nb = 1, 2, 3 \\
\hat{T} = \oplus_{nb=1}^3 \hat{T}_{nb}
\end{gather}

Here, $\hat{T}_{nb} \in \mathbb{R}^{(N \times (T \times th))} $ , and $\oplus$ represents concatenation,  thus, the outputs are concatenated to get $\hat{T} \in  \mathbb{R}^{(N \times 3 \times (T \times th))}$ . The application of  the above attention mechanism along the time dimension enables the identification of important time points. By focusing on these crucial time points, we can uncover temporal patterns that contribute to downstream classification tasks.

\subsection{Spatial Self-Attention Block}

Each brain ROI can be considered as a node in the graph, and the FC between ROIs shows how brain regions are linked with each other. To capture the directed FC between brain ROIs, we use a spatial self-attention module.

For each node $n \in N$, a sequence of $\hat{T}$ time point vector, denoted as $\hat{t}_n \in \mathbb{R}^T$, is fed into the self-attention module to compute the attention matrix $Att_{FC} \in \mathbb{R}^{N \times N}$ along with multi-head attention. In a simpler term, we're evaluating how each node relates to others over a sequence of time points. The following set of equations summarizes the process, where $\top$ represents transpose and $\oplus$ represents concatenation.

\begin{align}
k_n &= \hat{t}_n^\top W(k), \quad q_n = \hat{t}_n^\top W(q) \\\
K_A &= \oplus_{n=1}^{N} k_n, \quad Q_A = \oplus_{n=1}^{N} q_n \\\
Att_{FC} &= \text{softmax}(Q_AK_A^\top/\sqrt{d_{K_A}})
\end{align}

Here, $k_n$ and $q_n$ represent the key and query embeddings respectively for each input $\hat{t}_n$. The matrices $K$ and $Q$ are formed by concatenating all key and query embeddings across all nodes. 
% The attention matrix $W$ represents the attention weights calculated using softmax applied to the scaled dot product of query and key embeddings.
Here, the resulting $Att_{FC} \in \mathbb{R}^{N \times N}$ is the connectivity matrix between $N$ nodes in the graph, learned FC matrices for downstream classification. This enforces the model to estimate connectivity differences between classification groups (e.g., male and female or healthy controls and patients). The Directed Connectivity (DC)  estimated by the model represents  an interpretable graph to provide insights into task-dependent nodes and their connectivity.

\subsection{Graph Neural Network Block }
We make use of Relational Graph Isomorphism Network (RGIN) along with TopK Pooling to extract the spatial information from rs-fMRI data, which are explained further below.

\subsubsection{Constructing a connectivity graph }
The brain is spatially divided into N ROIs, which represent graph nodes indexed by the set \begin{math} V = 1,..., N \end{math}. The input node features are represented as a matrix \begin{math} H = [h_1,.....,h_N]^T\end{math} where \begin{math} h_i\end{math} is the feature vector of the node \begin{math} V_i\end{math}. An edge set \begin{math} E\end{math} represents the functional connections between ROIs with each edge  \begin{math} e_{ij} \end{math} linking two nodes \begin{math} (i,j) \in E \end{math}. The adjacency matrix $A_k \in R^{N \times N}$ is computed is formed by thresholded element   $e_{i,j} $  of correlation matrix computes as  Pearson's correlation coefficient between nodes' feature representation to achieve either fully connected or sparse graph. correlation matrix is computed  in two different ways: 
\begin{enumerate}
    \item 
 Pearson's correlation coefficient between nodes whose feature representation is the output of shared attention block with dimension  $\mathbb{R}^{(N \times 3 \times (T \times th))}$, which is dynamic in nature and changes during training, \item the Pearson's correlation coefficient between nodes whose feature representation is the original fMRI data series with dimension  $\mathbb{R}^{N \times T}$, constant throughout the training. 

 \end{enumerate}
The input node features,  \begin{math} h_i\end{math} $\in  \mathbb{R}^{1 \times N}$ , are the output of the self attention block which represents the directed connectivity strengths between node i to all other nodes.

The resultant undirected graph is denoted by the tuple \begin{math} G = (V, E) \end{math}.  
Let  \begin{math} (H = Att_{FC}) \in R^{N*N}, E \in R^{|E|*1} \end{math}, and  \begin{math} A \in R^{N*N} \end{math} symbolize the nodes features, edge features, and adjacency for the graph structure G.\\ 

\subsubsection{Relational Graph Isomorphism Network (RGIN)}

The RGIN convolution, introduced in our recent work \cite{BrainRGIN}, offers a novel approach to effectively capture both node and edge features in a graph by merging the concepts of Relational Graph Convolutional Networks (RGCN) and Graph Isomorphism Network (GIN). This method is particularly tailored for understanding brain function, where edge relations signify distinct functional clusters within the brain network. In simpler terms, nodes in the brain are categorized into functional clusters, and different clusters give rise to unique types of edge relations.

Unlike conventional approaches that fix cluster assignments for ROIs, the proposed RGIN model autonomously learns cluster formations. This means that nodes within the same cluster collaborate to achieve optimal performance for the given task. The integration of GIN aids in learning node embeddings for enhanced graph discriminative expression.

In RGIN, the aggregation function of GIN is replaced with that of RGCN, allowing the model to learn distinct mechanisms for different clusters. This design choice is motivated by the fact that GIN employs Multilayer Perceptrons (MLPs) to adapt to the complexity of brain connectivity, while RGCN excels in modeling the sub-network clustering nature of brain connectivity.

The forward propagation function of RGIN is defined as follows:

\begin{equation}
h_i^{(l)} = \text{MLP}^{(l)} \left( (1+\epsilon^{(l)})\cdot W_i^{(l)} \cdot h_i^{(l-1)} + \sum_{j\in N^{(l)}_{(i)}} W_j^{(l)} \cdot e_{i,j}^{(l-1)} \cdot h_j^{(l-1)}\right)
\end{equation}

In standard GNN models, $ W_i $ and $W_j $ are learnable weight matrices. However, in RGIN, they are defined as functions with parameters determined by the cluster assignment of nodes, making it an ROI-aware convolution layer. The model is trained end-to-end with a 2-layered MLP. The parameter $\epsilon^{(l)}$ is learnable and influences the importance of a node compared to its neighbors.

\begin{equation}
\textbf{W}^{(l)}_i  = \theta_2^{(l)} \cdot \text{relu}(\theta_1^{(l)} \textbf{r}_i + \textbf{b}^{(l)} )
\end{equation}
where ${W}^{(l)}_i$ is a function of the position encoding $r_i$, with  parameters $\theta_1$ and $\theta_2$ to build a output of dimension $d(l+1) \times d(l)$. $\text{ReLU}$ is the rectified linear unit activation function and $\textbf{b}^{(l)}$ is the bias term in the MLP. We represent each node's location information, $\textbf{r}_i$, using one-hot encoding instead of coordinates, assuming that the ROIs are aligned in the same order for all the brain graphs. 

Additionally, we assume that $\theta^{(l)}_1 = [\alpha^{(l)}_1, \ldots, \alpha^{(l)}_{N^{(l)}}]$, where $N^{(l)}$ is the number of ROIs in the $l^{th}$ layer, $\alpha^{(l)}_i = [\alpha^{(l)}_{i1}, \ldots, \alpha^{(l)}_{iK(l)}]^{T}$, $\forall i \in \{1, \ldots, N^{(l)}\}$, and $K(l)$ is the number of clusters. In this study, $K$ is selected as 7. $\alpha^{(l)}_i$ is the non-negative assignment scores of ROI $i$ to clusters. Assume $\mathbf{W}^{(l)} = [\beta^{(l)}_1, \ldots, \beta^{(l)}_{K(l)}]$ with $\beta^{(l)}_u \in \mathbb{R}^{d(l+1) \times d(l)}$, $\forall u \in \{1, \ldots, K(l)\}$, where $\beta^{(l)}_u$ is a basis matrix. Then, we can rewrite equation (4) as:

\begin{equation}
\textbf{W}^{(l)}_i = \sum_{u=1}^{K(l)} {\alpha}_{iu}^{(l)}  \beta^{(l)}_u + \textbf{b}^{(l)} 
% \textbf{W}^{(l-1)}_j = \sum_{u=1}^{K(l-1)} {\alpha}_{ju}^{(l-1)}  \beta^{(l-1)}_u + \textbf{b}^{(l-1)} 
\end{equation}

The incorporation of position encoding $r_i$ in ${W}^{(l)}_i$ is defined using parameters $\theta_1$ and $\theta_2$ to make the convolution layer specific to clustering. The final forward propagation function of RGIN Convolution can be expressed as:

\begin{equation}
\begin{split}
h_i^{(l)} = \text{MLP}^{(l)} \Biggl( &(1+\epsilon^{(l)}) \cdot \left(\sum_{u=1}^{K(l)} \alpha_{iu}^{(l)} \beta^{(l)}_u + \textbf{b}^{(l)}\right) \cdot h_i^{(l-1)} \\
&+ \sum_{j\in N^{(l)}_{(i)}} h_j^{(l-1)} \cdot \left(\sum_{u=1}^{K(l)} \alpha_{ju}^{(l)} \beta^{(l)}_u + \textbf{b}^{(l)} \right) \cdot e_{i,j}^{(l-1)} \Biggr)
\end{split}
\end{equation}
This formulation reduces the number of learnable parameters while still allowing a separate embedding kernel for each ROI, contributing to the model's efficiency.

\subsubsection{Pooling Layers}
 Via RGIN convolution, node-wise representations are generated. However, for the prediction task discussed in this paper, which requires graph-level prediction instead of node-level, these node-wise representations must be collected or pooled together. We apply ROI-aware TopK Pooling as it improves interpretability for rsfMRI data, by keeping the most indicative ROIs and removing the noisy and uninformative nodes (\cite{Li2021}), along with enforcing sparsity on the network. 
To select the most indicative ROIs, the choice of which nodes to drop is determined based on the scores of nodes obtained by projecting the node features to 1D via a learnable projection vector $\omega^{(l)} \in \mathbb{R}^{d^{(l)}}$. The pooled graph $(V^{(l+1)}, E^{(l+1)})$ is computed as follows:

\begin{enumerate}
  \item Calculate the scores of nodes $s^{(l)}$ with node feature matrix as follows:

   \begin{align}
   s^{(l)} = \frac{{H}^{(l)} w^{(l)}}{\|w^{(l)}\|_2}
   \end{align}

  \item Normalize the score $s^{(l)}$ by subtracting its mean and dividing by its standard deviation, yielding final scores $\tilde{s}^{(l)}$:

 \begin{align}
   \tilde{s}^{(l)} = \frac{(s^{(l)} - \mu(s^{(l)}))}{\sigma(s^{(l)})}
   \end{align}

  \item Find the top $k$ elements in the normalized score vector $\tilde{s}^{(l)}$ with $k$- largest values and get their indices as $i$:

 \begin{align}
   i = \text{topk}(\tilde{s}^{(l)}, k)
    \end{align}

  \item Create the pooled node features $H^{(l+1)}$ by element-wise multiplying the original node features $H^{(l)}$ with the sigmoid of the normalized scores and selecting only the elements indexed by $i$:

 \begin{align}
   H^{(l+1)} = (H^{(l)} \odot \text{sigmoid}(\tilde{s}^{(l)}))_{i:}
\end{align}

  \item Create new adjacency matrix as the new pooled graph $E^{(l+1)}$ by returning the edges between selected nodes $i$ from previous edge matrix $E^{(l)}$

   \[
   E^{(l+1)} = E^{(l)}_{i,i}
   \]
\end{enumerate}

In this description, $\|\cdot\|_2$ represents the $L_2$ norm, $\mu$ and $\sigma$ are functions that calculate the mean and standard deviation of a vector, \textit{topK} identifies the indices of the largest $k$ elements in a vector, $\odot$ denotes element-wise multiplication, and $(\cdot)_{i,:}$ selects elements in the $i^{th}$ row and all columns. \\

\subsubsection{Readout}
Previous studies have proven the effectiveness of the mean and max element-wise pooling operation(\cite{Li2021}). For a network, $G$ with $l$ convolution and respective pooling layers, the output graph of the $l^{th}$ pooling block is summarized using a mean and max pooling operation element-wise on $H(l)=[h(l)_i:i=1,...,N(l)]$. The resulting vector is obtained by concatenating both the mean and max summaries. To obtain a graph-level representation, the summary vectors from each layer of the RGIN block are concatenated together, and a 2-layered MLP is used for the final classification.\\

\section{Experiments Setup and Model Overview}

\label{sec:replication_section}
\subsection{Model Overview}

Our model as presented in Fig. \ref{DSAM_Architecture} begins with the utilization of a TCN architecture to extract the temporal dynamics of rs-fMRI time series for each node in the network. %This process unfolds through the incorporation of three blocks, each with a pair of 1D convolutional layers along with 1D batch normalization, ReLU activation, and dropout. 
Within the three blocks, a kernel of size 7, denoted as $K=7$, is employed, and the number of channels is 8, 16, and 32 respectively. The dilation factor $d$ adopts the form $d=2^{(l-1)}$, where $l$ signifies the block (i.e., $l\in\{1,2,3\}$). 
% Consequently, the TCN model yields an output shape of $\mathbb{R}^{(N \times 3 \times (T \times th))}$ after all three blocks, where $\textit{$\hat{t_i}$}$ for each node is 1200.

Following this, the model transitions into a Temporal Attention Block featuring a Transformer-based \textit{Shared Attention} block \cite{TransformerPaper}. This block incorporates a single-layer attention component with $\textit{number\_heads}=8$. After extracting the attention scores, we did a hyperparameter search to finalize the optimal threshold for selecting the top-$k$ time points ($th$) from the provided range of values: $[1,0.8,0.6,0.4, 0.2, 0.1]$, which resulted in 0.1 providing enough informative feature representations to be used further. The resultant feature set assumes dimensions of $\mathbb{R}^{(N \times 3 \times (T \times th))}$. The $\textit{th}$ as 0.1 or just 10\% significantly reduces the time points (using only 360 time points from a total of 3600 time points from three blocks of TCN) and thus the total parameter count later on. 

Subsequently, this feature set is passed to the self-attention block to calculate the learned connectivity matrix. The self-attention Block comprises W matrices for key, value, and query operations with the embedding size of 64. The number of attention heads, denoted as $\textit{num\_attention\_heads}$, is fixed at 8. The resulting learned connectivity matrix emerges as a node feature representation with dimensions of $\mathbb{R}^{1\times N}$ for each node, which subsequently serves as input to an RGIN model.

Adjacency matrix within the graph structure is constructed using one of two methods:   dynamic edges from TCN outputs,  and constant edges from original time series.  As a result of the hyperparameter search, a fixed threshold of 30\% is applied to build a sparse edge representation for both dynamic and constant edge scenarios. We used two layers of RGIN convolution block of sizes [32,32] with a pooling ratio of 0.5, and a 2-layered fully connected layer with sizes [32, 512] with ReLU activation functions in between. The number
of clustered communities for RGIN block is set to 7, and the motivation for this comes from the seven functional networks defined by (\cite{ThomasYeo2011})  to show key brain functional domains.  In the final step, the output from the RGIN  feeds into a linear model for the classification task.

\subsection{Training the Model}

The DSAM architecture was implemented using Pytorch \cite{PazkeTorch}, and Pytorch Geometric \cite{TorchGeometric} for the specific graph neural network components. The number of nodes was 100 (corresponding to Schaefer Atlas), and the number of node features per node was the number of time points for each node (i.e., 1200). We tested different versions of the architecture with dynamic or constant edges. To evaluate the performance and validity of our model, we conducted proof-of-concept experiments using a widely recognized binary sex prediction task \cite{Jiang2019, Weis2019}. Our approach employed a 5-fold stratified cross-validation procedure on the HCP dataset. In each fold, we divided the dataset into training and test sets, with the test set comprising 20\% of the original data.

The neural network training of our model was performed for 150 epochs, utilizing the Adam optimizer \cite{Adam} and the Cross-Entropy loss function \cite{CrossEntropy}. An early stopping mechanism was incorporated, terminating training if the validation loss failed to decrease over 40 consecutive epochs. Additionally, the learning rate was subject to reduction by a factor of 0.1 with a patience setting of 30. We utilized a batch size of 32. Within each random run, we preserved the model with the smallest validation loss. The model exhibiting the lowest validation loss was evaluated on the respective test set. This process was independently executed for each of the five test sets. Metrics derived from these evaluations were subsequently averaged to produce the final results.

We also compared the performance of \technique \textbf{}{ } with popular recent models along with traditional robust ML models as baseline models. The recent baselines includes BrainGNN (\cite{Li2021}), BrainRGIN (\cite{BrainRGIN}), BrainNetCNN (\cite{BrainNetCNN}), Brain Network Transformer \cite{BrainNetworkTransformer}, FBNetGen \cite{FBNetGen}, Graph Transformers \cite{GraphTransformer} etc. Furthermore, we also compared the performance of our model with a stable deep spatiotemporal model \cite{Azevedo2022} that also uses TCNs as the initial block to capture the temporal information, but afterward makes use of node and edge models to capture the spatial dependencies. Furthermore, we compared with the baseline models like GAT, and SVM. Based on performance, setting the learning rate of both BrainRGIN and BrainGNN initially  to 0.001 and reducing every 30 epochs achieved the best performance for all prediction tasks. For BrainGNN, we used two convolutional layers of size 32, as presented in the original paper (\cite{Li2021}), with a fully connected layer of size 512. We utilized the authors' provided open-source codes for BrainGNN \cite{Li2021}, FBNetGen \cite{FBNetGen} and Brain Network Transformer (BNT) \cite{BrainNetworkTransformer}, conducting grid searches to fine-tune essential hyperparameters based on the recommended best configurations. Specifically, for BrainGNN, we also explored various learning rates (0.01, 0.005, 0.001) while for FBNetGen, we investigated how varying hidden dimensions (8, 12, 16) impact the performance. For BNT, we used different hidden dimensions (256, 512, 1024) during scaled dot product attention, along with varying fully connected layers (2 and 3).  In the case of BrainNetCNN \cite{BrainNetCNN}, we experimented with different dropout rates (0.3, 0.5, 0.7). Regarding GT (Graph Transformer) \cite{GraphTransformer}, we examined the impact of different number of transformer layers (1, 2, 4) and the number of attention heads (2, 4, 8) over 50 epochs of training. For the training of the model by \cite{Azevedo2022}, we used a hyperparameter sweep referencing the parameters suggested in the original paper. The input feature used for this model was also the direct rs-fMRI time points. The hyperparameters used were: batch size = 32, optimizers = [Adam, Rmsprop], dropout a range of values between 0 and 0.5, number of nodemodel layers = [2,3], temporal embedding dimension = [16,32], lr a range of values between 0.01 and 0.00001. The learning rate was scheduled to reduce by a factor of 0.1, same as above with a patience of 30. Other experimental settings were fixed as in the original paper \cite{Azevedo2022}. Besides, we experimented with different numbers of heads = [2,4,6] for GAT, where we also tested with FC features as input features. For SVM, we used Polyssifier (https://github.com/sergeyplis/polyssifier). This tool is widely used to perform baseline model comparisons and has a feature to automatically perform hyperparameter search and identify the best parameters of the model for the highest performance. The input features for SVM was directly the cross-correlation between each of the nodes, which is termed the Functional Connectivity (FC) features.

\section{Results}
\label{sec:result}
In this section, we present the results of our experiments comparing DSAM with the baseline models, and evaluating different forms of the DSAM architecture with varying components. We aim to assess the impact of different architectural choices on the performance of DSAM in capturing dynamic connectivity patterns in brain networks.

\subsection{Comparison with Baseline Models}

Table~\ref{tab:baseline-comparision} provides a comparison of DSAM with baseline models in terms of various evaluation metrics. We report the number of edges, Area Under the ROC Curve (AUC), accuracy, sensitivity, and specificity.

% \begin{table*}[]
% \caption{Comparison with baseline models on HCP Dataset}
% \label{tab:baseline-comparision}
% \tiny
% \resizebox{\linewidth}{!}{%
% \begin{tabular}{|l|l|l|l|l|l|}
% \hline
% \textbf{Model} & \textbf{AUC} & \textbf{Accuracy} & \textbf{Sensitivity} & \textbf{Specificity} \\ \hline
% DSAM (Ours)- Dynamic Edges & \textbf{83.02 \textpm 1.5} & \textbf{82.05 \textpm 0.83} & 85 \textpm 0.8 & \textbf{85.18 \textpm 0.5} \\ \hline
% DSAM (Ours) - Constant Edges & 81.12\textpm 1.7 & 80.06 \textpm 0.4 & 84.12\textpm 1.3 & 84.34 \textpm 0.8 \\ \hline
% Brain Network Transformer \cite{BrainNetworkTransformer}  & 82.82\textpm 0.7 & 81.5 \textpm 0.8 & \textbf{85.06\textpm 1.7} & 85.34 \textpm 0.4 \\ \hline
% BrainNetCNN \cite{BrainNetCNN} & 74.3\textpm 2.31 & 75.21 \textpm 2.1 & 74.12\textpm 2.7 & 77.34 \textpm 1.2 \\ \hline
% FBNetGen  & 80.12\textpm 1.8 & 81.12 \textpm 1.8 & 83.75\textpm 1.3 & 79.34 \textpm 0.9 \\ \hline
% BrainRGIN \cite{BrainRGIN} & 77.51\textpm 0.9 & 79.96 \textpm 2.5 & 84.12\textpm 1.3 & 74.34 \textpm 0.2 \\ \hline
% N+E (Concat) (\cite{Azevedo2022}) & 77.12 & 77.01 & 79.32 & 80.16 \\ \hline
% GAT (\cite{GAT}) & 78.12 \pm 3.34 & 77.13 \pm 2.7 & 68.12 & 69.32 \\ \hline
% SVM  & 75.32 & 73.21 & 72.56 & 74.34 \\ \hline
% % XGBoost & None & 72.01 & 76.03 & 74.56 & 72.32 \\ \hline
% \end{tabular}
% }
% \end{table*} 

\begin{table*}[]
\caption{Comparison with baseline models on HCP Dataset}
\label{tab:baseline-comparision}
\tiny % Changed from \tiny to \small for better readability
\resizebox{\linewidth}{!}{%
\begin{tabular}{|l|l|l|l|l|}
\hline
\textbf{Model} & \textbf{AUC} & \textbf{Accuracy} & \textbf{Sensitivity} & \textbf{Specificity} \\ \hline
DSAM (Ours) - Dynamic Edges & \textbf{$83.02 \pm 1.5$} & \textbf{$82.05 \pm 0.83$} & $85 \pm 0.8$ & \textbf{$85.18 \pm 0.5$} \\ \hline
DSAM (Ours) - Constant Edges & $81.12 \pm 1.7$ & $80.06 \pm 0.4$ & $84.12 \pm 1.3$ & $84.34 \pm 0.8$ \\ \hline
Brain Network Transformer \cite{BrainNetworkTransformer} & $82.82 \pm 0.7$ & $81.5 \pm 0.8$ & \textbf{$85.06 \pm 1.7$} & $85.34 \pm 0.4$ \\ \hline
BrainNetCNN \cite{BrainNetCNN} & $74.3 \pm 2.31$ & $75.21 \pm 2.1$ & $74.12 \pm 2.7$ & $77.34 \pm 1.2$ \\ \hline
FBNetGen & $80.12 \pm 1.8$ & $81.12 \pm 1.8$ & $83.75 \pm 1.3$ & $79.34 \pm 0.9$ \\ \hline
BrainRGIN \cite{BrainRGIN} & $77.51 \pm 0.9$ & $79.96 \pm 2.5$ & $84.12 \pm 1.3$ & $74.34 \pm 0.2$ \\ \hline
N+E (Concat) \cite{Azevedo2022} & $77.12$ & $77.01$ & $79.32$ & $80.16$ \\ \hline
GAT \cite{GAT} & $78.12 \pm 3.34$ & $77.13 \pm 2.7$ & $68.12$ & $69.32$ \\ \hline
SVM & $75.32$ & $73.21$ & $72.56$ & $74.34$ \\ \hline
% XGBoost & None & 72.01 & 76.03 & 74.56 & 72.32 \\ \hline
\end{tabular}
}
\end{table*}

Our DSAM architecture, both in dynamic and constant settings, outperforms the baseline models across all metrics. In the dynamic setting, DSAM achieves an AUC of 83.02\%, an accuracy of 82.05\%, and strong sensitivity and specificity scores of 85\% and 86\%, respectively. Even in a constant setting, DSAM maintains competitive performance with an AUC of 81.12\% and an accuracy of 80.06\%. The comparison with the recent model by \cite{Azevedo2022} baseline reveals significant improvement over existing methods that do not learn the connectivity matrix. 
Furthermore, we observed stable and comparable performances from baseline models such as BrainNetCNN \cite{BrainNetCNN}, FBNetGen \cite{FBNetGen}, and GT \cite{GraphTransformer}, indicating the reliability of these models. However, the Brain Network Transformer (BNT) \cite{BrainNetworkTransformer} exhibited robust performance with stable metrics. These results demonstrate the effectiveness of DSAM in capturing and utilizing dynamic connectivity patterns in brain networks, providing superior classification performance compared to traditional baseline models.

\subsection{Experiment with the Adolescent Brain Cognitive Development (ABCD) Dataset}

\subsubsection{Hyperparameter settings for ABCD Dataset}
Similar to the HCP dataset, we performed a binary sex prediction task to evaluate the performance and validity of our model.  The configuration the model is  same, while the time points of input node was 360.  We performed a hyperparameter search to finalize the optimal threshold for selecting the top-k time points (tk) from the provided range of values:[1, 0.8, 0.6, 0.4], similar to HCP. We employed a 5-fold cross-validation procedure. All other hyperparameter settings remain the same.

\subsubsection{Results from ABCD Dataset}

In the context of ABCD, our architecture \technique \text{ } outperforms most of the baseline models as can be seen in Table \ref{tab:baseline-comparision-abcd}. In the dynamic setting, DSAM achieves an AUC of 84.12\%, an accuracy of 83.27\%, an AUC of 81.12\%, and an accuracy of 82.06\% in the constant setting. Similar to the experiment with HCP datasets, baseline models such as BrainNetCNN (\cite{BrainNetCNN}), FBNetGen (\cite{FBNetGen}), Brain Network Transformer (BNT) (\cite{BrainNetworkTransformer}), and GT (\cite{GraphTransformer}) exhibited stable and comparable performances indicating the reliability of these models in the analysis.

\begin{table*}[]
\caption{Comparison with baseline models on ABCD Dataset}
\label{tab:baseline-comparision-abcd}
\tiny
\resizebox{\linewidth}{!}{%
\begin{tabular}{|l|l|l|l|l|l|}
\hline
\textbf{Model} & \textbf{AUC} & \textbf{Accuracy} & \textbf{Sensitivity} & \textbf{Specificity} \\ \hline
DSAM (Ours)- Dynamic Edges  & \textbf{85.12 \textpm 1.5} & 83.27 \textpm 1.75 & 85 \textpm 1.8 & 86 \textpm 0.9 \\ \hline
DSAM (Ours)- Constant Edges & 81.12\textpm 1.2 & 82.06 \textpm 1.1 & 84.12\textpm 1.3 & 84.34 \textpm 0.9 \\ \hline
Brain Network Transformer \cite{BrainNetworkTransformer}  & 84.98\textpm 2.1 & \textbf{84.12} \textpm 1.73 & \textbf{86.32\textpm 1.9} & \textbf{87.12 \textpm 1.2} \\ \hline
BrainNetCNN \cite{BrainNetCNN} & 78.21\textpm 1.3 & 79.22 \textpm 1.8 & 78.23\textpm 1.3 & 80.43 \textpm 1.1 \\ \hline
FBNetGen  & 82.12\textpm 2.1 & 82.90 \textpm 1.4 & 84.91\textpm 1.1 & 82.79 \textpm 0.7 \\ \hline
BrainRGIN \cite{BrainRGIN} & 78.23\textpm 1.6 & 80.19 \textpm 2.5 & 81.48\textpm 0.8 & 83.19 \textpm 0.6 \\ \hline
N+E (Concat) (\cite{Azevedo2022}) & 79.26 & 80.92 & 82.13 & 83.66 \\ \hline
\end{tabular}%
}
\end{table*} 

\subsection{Ablation Studies}

\subsubsection{Question 1: How important are each blocks of \technique \textbf{ } architecture ?}

\begin{table}[htbp]
\caption{Ablation studies with different forms of DSAM architecture on HCP dataset}
\label{tab:ablation-studies}
\tiny
\resizebox{\columnwidth}{!}{%
\begin{tabular}{|l|l|l|l|l|}
\hline
\textbf{Architecture} & \textbf{Edges} & \textbf{GCN Features} & \textbf{AUC} & \textbf{Accuracy} \\ \hline
\begin{tabular}[c]{@{}l@{}}DSAM (TCN + Temporal\\ Attention+Self Attention+RGIN)\end{tabular} & Dynamic & Learned FC& \textbf{83.02} & \textbf{82.05} \\ \hline
\begin{tabular}[c]{@{}l@{}}TCN+Self Attention+RGIN\end{tabular} & Dynamic & Learned FC& 79.52 & 79.05 \\ \hline
\begin{tabular}[c]{@{}l@{}}TCN+Temporal Attention+\\ RGIN\end{tabular} & Dynamic & TCN outputs & 73.29 & 73.98 \\ \hline
\begin{tabular}[c]{@{}l@{}}TCN + Temporal Attention+\\ Self Attention + 2 Layered MLP\end{tabular} & - & - & 63.29 & 66.98 \\ \hline
RGIN (Only GNN) & Constant & Static FC& 77.51 & 79.96 \\ \hline
\end{tabular}%
}
\end{table}

Our ablation studies (performed on the HCP dataset), detailed in Table~\ref{tab:ablation-studies}, shed light on the nuanced relationships among various components within the DSAM architecture and their influence on performance metrics, particularly AUC and accuracy. The comprehensive DSAM architecture, encompassing TCNs, Temporal Attention, Self Attention, and GNNs, exhibits notable efficacy with an AUC of 83.02\% and an accuracy of 82.05\% in dynamic scenarios.  When specific components are selectively removed or modified, a discernible decline in performance is observed. For instance, the exclusion of Temporal Attention resulted in a drop of accuracy to 79.52\% which as a similar case with 100\% time points and with a tremendous amount of timepoints features in the self-attention block with huge performance load. Similarly, leaving out the Self-Attention component and directly using the output of Temporal Attention as input features to GNN, we saw a notable drop in AUC to 73.29\% and accuracy to 73.98\%. This underscores the pivotal role played by Temporal and Self-attention in extracting essential temporal features and refining the model's understanding of complex relationships in dynamic settings. Furthermore, the role of GNN (RGIN) in capturing the spatial dynamics is also vividly evident from Table 3, where dropping GNN and using 2-layered MLP directly for prediction after Self-Attention resulted in a significant drop in accuracy.

In summary, our ablation studies highlight the intricate interplay among TCN, Temporal Attention, Self Attention, and GCN within the DSAM architecture. The findings underscore the necessity for a holistic integration of these components to effectively capture dynamic connectivity patterns, showcasing the significance of each element in maintaining the model's overall performance integrity. These insights provide valuable guidance for refining and optimizing the DSAM architecture for applications involving dynamic learning scenarios.

\subsubsection{Question 2: What percentage of original time points are informative for the prediction?}

In our investigation of the HCP dataset, we aimed to understand how the selection of time points influences our model's predictive performance. Through a hyperparameter search across the threshold ($\textit{th}$) ranges of 10\%, 20\%, 40\%, 60\%, 80\%, and 100\%, we made a surprising discovery: our model achieved its optimal performance with just 10\% of the original time points. This selection dramatically reduces the number of features processed by the subsequent self-attention block, with only 120-time points retained per level of TCN from 1200 time points. Considering three levels of TCN, this amounts to a total of merely 360-time points utilized for prediction. Despite this reduction, these selected time points encapsulate crucial information, thanks to feature concatenation occurring at multiple levels within the model architecture. Notably, this reduction in the number of time points also corresponds to a decrease in the total number of parameters, further streamlining our model's computational load.

\begin{figure*}
\centering
\includegraphics[width=0.7\linewidth]{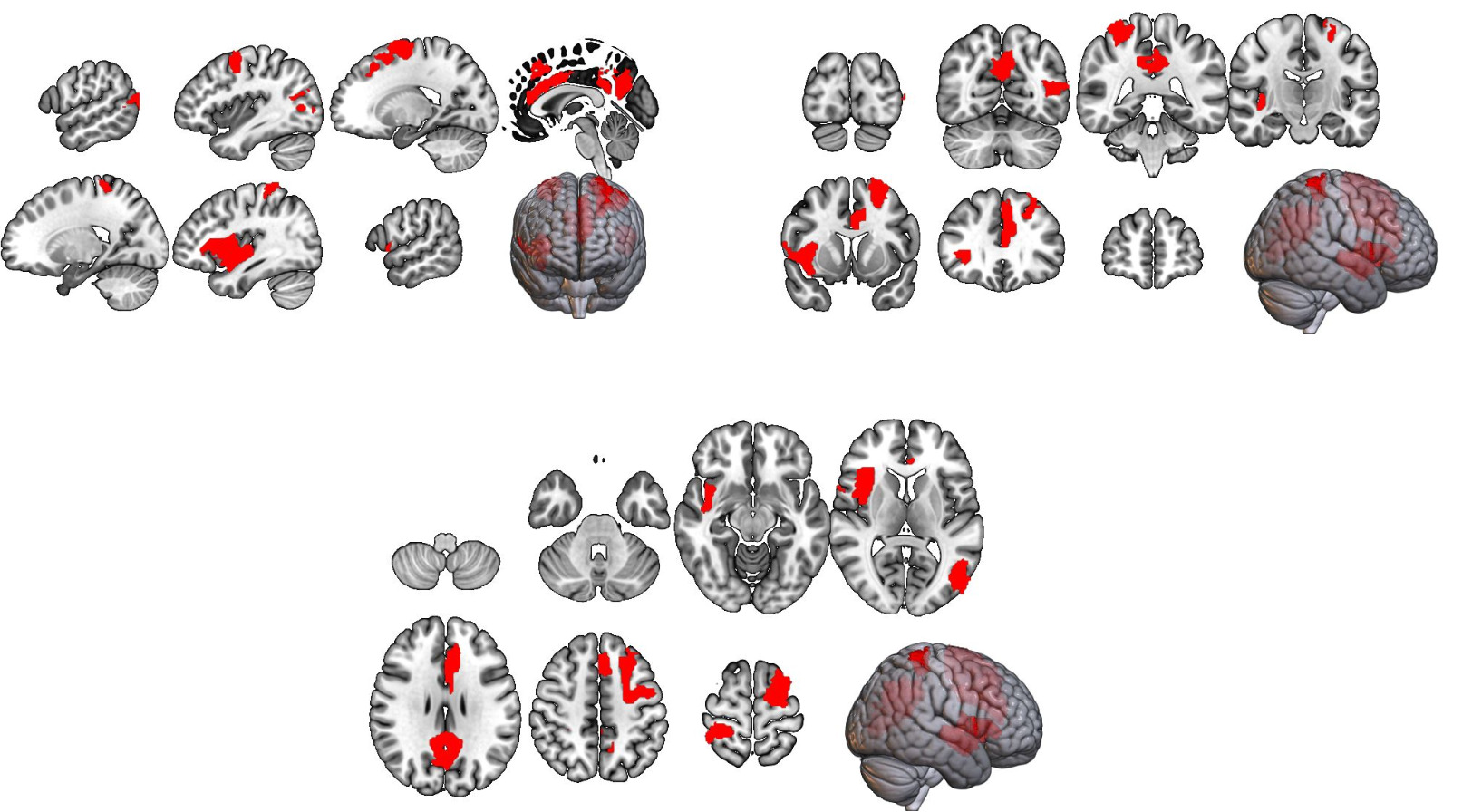}
\caption{ Regions of significant differences between male and female (HCP)}
\label{Significant_Node_Networks}
\end{figure*}

\subsubsection{Question 3: How does performance alter when we select a larger number of nodes?}

We also aim to understand how the number of ROI selections impacts the performance of the model, and if the model learns better with more nodes on the HCP dataset. For this, we used the same atlas \cite{Schaefer2017} atlas for brain parcellation, but with 200 and 400 regions. As this drastically increases the parameters of the model, we reperformed the hyperparameter optimization for just the dropout selecting a different number of parameters, where dropout of 0.65 for 200 regions and 0.68 for 400 regions gave the best results of 82.15\% and 83.01\% accuracy. The results reveal that the results are pretty similar, and when optimized, reveal very similar performance.

\subsubsection{Question 4: How many blocks of TCN should be used for the best results??}

In this section, we answer the idea behind selecting 3 blocks of TCN. The aim of using multiple blocks in TCN is to have diverse representations such that the network can learn a more comprehensive and diverse set of features. Because the dilation increases exponentially, the first block captures low-level temporally local patterns, while subsequent blocks can learn more high-level, large-range, complex features. This hierarchy allows the network to understand the data at different levels of abstraction, potentially improving the model's ability to capture intricate patterns present in resting state fMRI data. Therefore, we experimented with different numbers of blocks (2, 3, and 4) to see what number of blocks in TCN presented us with the best results. Using the same model configurations, 2,3 and 4 blocks reported an accuracy of 79.12\%, 82.05\%, and 81.95\%, which is an evidence that three-level of feature dilation is enough to capture enough temporal patterns in rs-fMRI data. We did not experiment with blocks larger than 4 blocks because the complexity and parameter of the whole model increased significantly.

\begin{figure*}
  \centering
  \begin{subfigure}[b]{0.8\linewidth}
    \centering
    \includegraphics[width=\linewidth]{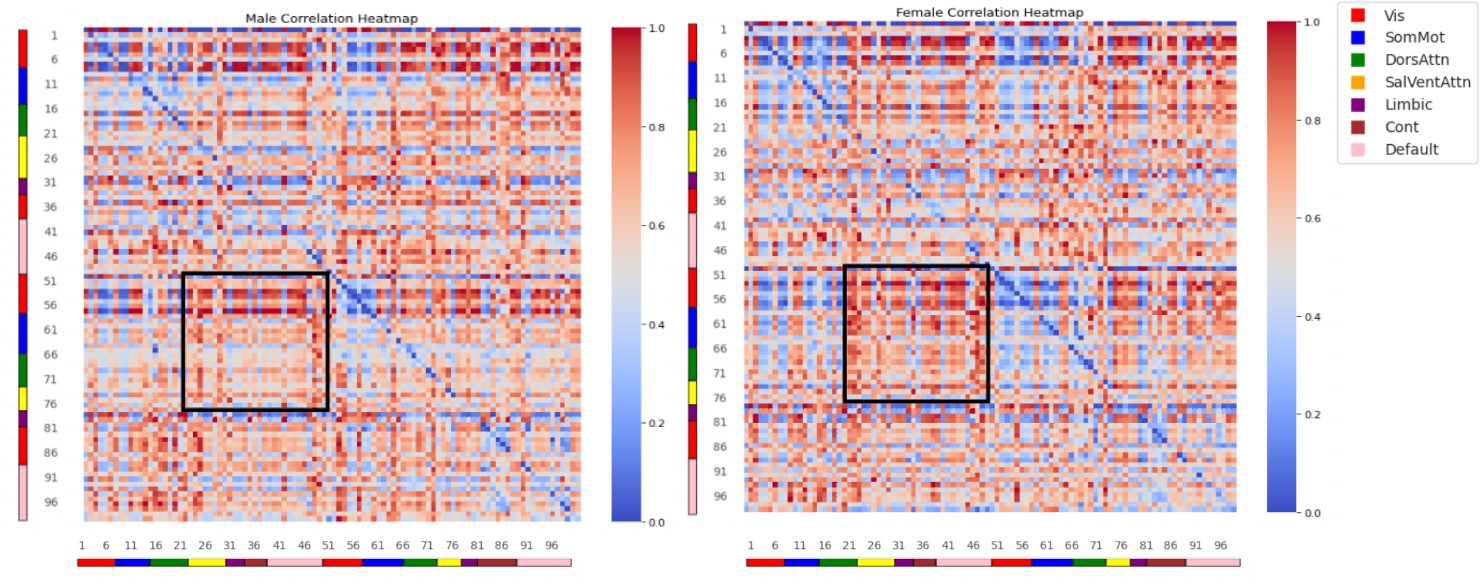}
    \caption{Comparison of Learned Directed Connectivity via Self-Attention for Males vs Females}
    \label{fig:Male_dfnc_corr}
  \end{subfigure}
  \hfill
  \begin{subfigure}[b]{0.8\linewidth} % Adjust the width as needed
    \centering
    \includegraphics[width=0.5\linewidth]{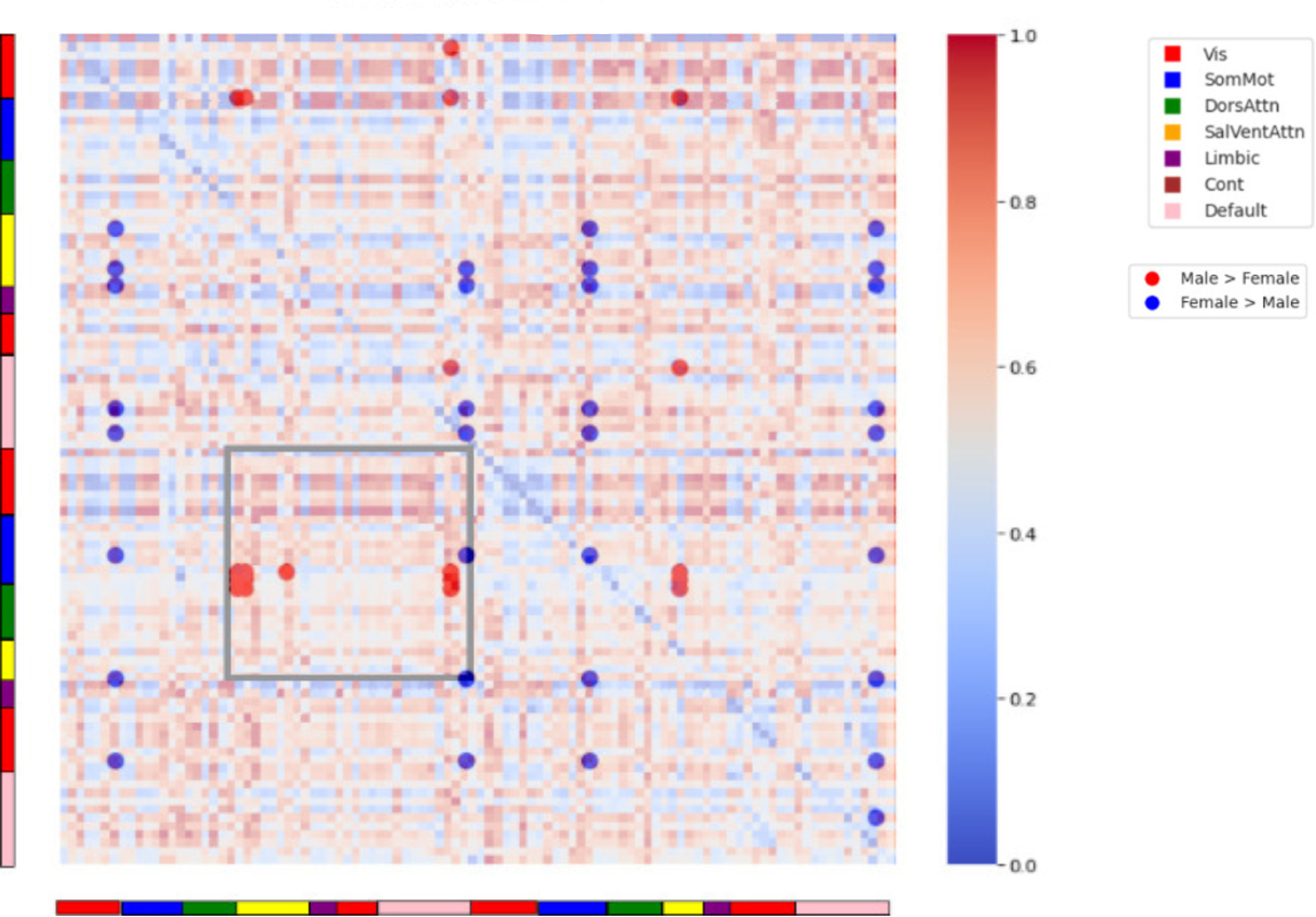}
    \caption{ Visualizing the Direction of Significance of Nodes in Relation to Sex}
    \label{fig:Direction_Significance}
  \end{subfigure}
  \hfill
  \caption{(a) Comparison of learned directed FC between males and females, highlighting regions of interest. (b) Visualization of the direction of significance of nodes concerning sex differences.}
  \label{fig:combined}
\end{figure*}

\subsection{Directed Functional Connectivity for Sex for HCP dataset}

Using the model's inbuilt interpretable power, we plot the learned functional connectivity matrix. The output of the trained self-attention block was extracted from a holdout test set and normalized along the column dimension. The male and female connectivity matrixes were extracted, and the correlation heatmaps are presented in Figure \ref{fig:combined}.  The 100 ROIs are organized into seven functional networks:  \textit{Visual network; Sensorimotor network; Dorsal attention network; Ventral attention network; Limbic network; Control network; Default mode network.}

After extracting the connectivity matrix for both males and females, we performed t-test on all pairs of connectivity between males and female to analyze the results. Multiple comparison with Bonferroni correction was done for both tests.

For the t-test between all pairs of 100 nodes, we found connectivity between 52 pairs of nodes (18 nodes involved ) to be significant (p-value $\leq 5\times 10^{-5}$). The most involved nodes included the right postcentral gyrus, left calcarine sulcus, left middle frontal gyrus, right and left posterior cingulate cortex (PCC), left cingulum bundle, and right insular cortex. These regions represent key components of the brain involved in somatosensory processing, visual perception, executive functions, default mode network activity, emotional regulation, and social cognition, which we elaborate further in discussion section below. Furthermore, upon grouping the nodes within the 7 connectivity networks, the default mode network had the highest number of nodes (n=5) to be sex differential, whereas the limbic network had zero significant nodes. The complete number of nodes per networks are represented in Figure \ref{Significant_Node_Networks}.

\begin{figure}
\centering
\includegraphics[width=0.7\columnwidth]{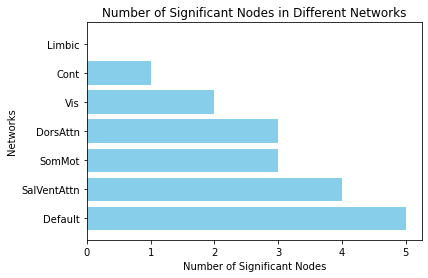}
\caption{ Significant nodes within each network (HCP)}
\label{Significant_Node_Networks}
\end{figure}

\section{Discussion and Conclusion}
\label{sec:dic_con}
%Ablation studies, and the interpretability of the manuscript.
In this research study, we have demonstrated the effectiveness of our model, \textit{\technique}, in capturing the spatiodynamics of the brain and learning connectivity matrices relevant to prediction tasks. Our model leverages TCN, temporal attention, self-attention, and ROI-aware GNNs to comprehensively capture the spatiotemporal dynamics of brain connectivity directly from raw and noisy resting-state fMRI (rs-fMRI) data. Significantly, this approach enhances interpretability and offers a deeper understanding of the mechanisms underlying brain function.

One of the key strengths of our model lies in its direct utilization of raw resting-state fMRI time points, bypassing the need for handcrafted FC features commonly employed by existing models such as BNT, FBNetGen, BrainRGIN, and BrainGNN. This approach represents a departure from conventional methods and presents a more challenging task of seamlessly integrating both temporal and spatial blocks, thus contributing to the model's "heaviness." Moreover, our model offers a significant advantage in its ability to learn directed goal-specific connectivity matrices. The directed FC matrices are different from correlation-based static FC (see the Supplementary figure for static FC), suggesting that the model is able to learn new patterns as well as learn effectively critical nodes and connections within the brain for the task. This capability holds immense promise for uncovering relevant phenotypes and advancing our understanding of brain function. Though having differences, we have some similarities in static FC vs learned directed FC for sex where in both t-tests, default mode had the highest number of significant nodes (23 nodes in static FC and 5 nodes in learned FC), and the limbic network had the lowest number of nodes (3 for static FC, 0 for learned FC). This comparison highlights the nuanced differences between static FC and our goal-oriented learned FC, underscoring the relevance and specificity of our approach in uncovering brain connectivity patterns tailored to the task at hand.

In this research study, our decision to focus more on the HCP dataset over the ABCD dataset is rooted in several factors. Firstly, the HCP dataset offers a sample size almost ten times smaller than the ABCD dataset. This smaller sample size makes it more manageable for training and running experiments efficiently, allowing us to explore our model's capabilities more effectively within the constraints of computational resources.

The results obtained from the interpretation perfectly align with existing research findings in the field of neuroscience. Specifically, research has consistently demonstrated differential activation of the right postcentral gyrus during mental rotation tasks, indicating potential sex-specific cognitive strategies or neural mechanisms \cite{Butler2006}. Furthermore, sex differences observed in regions like the Calcarine have been linked to affective disorders, reinforcing the significance of considering sex-specific neurobiological factors in understanding these conditions \cite{Chaudhary2023,Tu2022,Piani2022}. Similarly, investigations into middle frontal volume and connectivity have underscored associations with antisocial personality and autism, emphasizing the importance of sex-specific considerations in elucidating the neurobiological underpinnings of these disorders \cite{Piani2022,Raine2009}. Moreover, the role of the posterior cingulate cortex (PCC) within the Default Mode Network (DMN) has been highlighted, with findings suggesting significant sex-by-aging interaction and implications for conditions like autism and substance use disorders \cite{Scheinost2014,Jung2015,Ritchay2021}. Similarly, investigations into the insula and cingulum, components of the saliency network, have revealed sex differences in connectivity patterns, particularly in the context of autism spectrum disorders, further emphasizing the relevance of sex-specific neural mechanisms in understanding these conditions \cite{Cummings2020}. Additionally, sex differences in morphological measures, particularly those related to interoception, provide further insight into the complex interplay between brain structure, function, and the manifestation of cognitive and affective processes \cite{Longarzo2021}. Integrating findings from these diverse brain regions and networks not only enhances our understanding of sex-specific neural mechanisms but also holds promise for personalized approaches to diagnosis and treatment across various neurological and psychiatric conditions.

As we observe the difference in results when using dynamic edge weights as compared to static edge weights, we assume this is because of the graph being adapted with the temporal features over time, which extracts more relevant edge features as compared to using static edges. This difference helps to adjust the graph over time more specifically to the task at hand. As the main focus of the model is to learn goal-specific connectivity of the brain, we also believe having dynamic edges did open more flexibility for the model to be more oriented towards a specific task.

We want to point out the limitations of the model being "heavy" and performance-resource tradeoffs. Fully aware of its computational weights, each component of our model's architecture is carefully designed to satisfy the task at hand. That said, we also optimized the model for a lighter computation load by utilizing only a fraction of the original time points from the TCN.  Specifically, we retain just 10\% of the original time points, significantly reducing the computational load and parameter count, reflecting our deliberate efforts to streamline the analysis process while maintaining performance. Despite the computational demands, the performance gains achieved over simpler models warrant focused and merit attention. The decision to employ a simpler or heavier model depends on case-by-case careful consideration of performance-resource trade-offs.  Our purpose here is to present a novel model architecture meticulously designed to align with the requirements of the task at hand, modeling temporal-spatial dynamics of brain function for a specific task.  

In conclusion, our model represents a significant step forward in computational neuroscience, offering a powerful tool for analyzing and interpreting brain connectivity dynamics directly from raw rs-fMRI data.    As a preliminary version of this study, future work will include additional experiments on specific tasks (such as classifying patients with schizophrenia or Alzheimer's, predicting cognitive ability). Its utility and effectiveness opens up great potential to gain deeper insights into how the human brain adapts its functional connectivity specific to the task.

\section*{Data and Code availability for replication}

The code is openly available at \url{https://github.com/bishalth01/DSAM}. The important parameters for replication are mentioned above in Section \ref{sec:replication_section}, and also in readme.txt file inside the GitHub. Data cannot be open-sourced due to restrictions but can be provided upon special request.

\section*{Acknowledgments}
This study was funded part by NIH grant R01DA049238 and NSF grant 2112455. We thank the Adolescent Brain Cognitive Development (ABCD) participants and their families for their time and dedication to this project. Data used in the preparation of this article were obtained from the Adolescent Brain Cognitive Development (ABCD) Study (https://abcdstudy.org), held in the NIMH Data Archive (NDA). This is a multisite, longitudinal study designed to recruit more than 10,000 children aged 9-10 and follow them over 10 years into early adulthood. The ABCD Study is supported by the National Institutes of Health and additional federal partners under award numbers U01DA041048, U01DA050989, U01DA051016, U01DA041022, U01DA051018, U01DA051037, U01DA050987, U01DA041174, U01DA041106, U01DA041117, U01DA041028, U01DA041134, U01DA050988, U01DA051039, U01DA041156,
U01DA041025, U01DA041120, U01DA051038, U01DA041148, U01DA041093, U01DA041089, U24DA041123,
U24DA041147. A full list of supporters is available at https://abcdstudy.org/federal-partners.html. A listing of participating sites and a complete listing of the study investigators can be found at https://abcdstudy.org/consortium\_members. ABCD consortium investigators designed and implemented the study and/or provided data but did not necessarily participate in analysis or writing of this report. This manuscript reflects the views of the authors and may not reflect the opinions or views of the NIH or ABCD consortium investigators.

\section*{Author contributions statement}

B.T. conducted the primary data analysis, coding implementation, experiments, writing, and reviewing of the manuscript. R.M, Y.P.W, and E.A. actively participated in reviewing the design of the model and provided insightful suggestions while reviewing the manuscript. R.S, B.R., and P.S were involved in the in-depth discussion of the analyses and reviewing the manuscript. J.C. assisted in analyzing the results, and also assisted in interpretation. S.G contributed to the theoretical analysis of the overall architecture and participated in reviewing the manuscript.  V.D.C was involved in the design of the project and the manuscript review.  J.L. was involved in the design, data analysis, and revising the manuscript. All authors reviewed the manuscript. 

\section*{Financial Disclosures Section}

All authors declare that they have no conflicts of interest.

\bibliography{main_manuscript} 
\bibliographystyle{unsrtnat}

%%% Uncomment this line and comment out the ``thebibliography'' section below to use the external .bib file (using bibtex) .





% \end{thebibliography}

\end{document}